# All-Si Valley-Hall Photonic Topological Insulator

Tzuhsuan Ma and Gennady Shvets*
Department of Physics, The University of Texas at Austin, Austin, Texas 78712, USA
*gena@physics.utexas.edu

*Abstract:*
An all-Si photonic structure emulating the quantum-valley-Hall effect is proposed. We show that it acts as a photonic topological insulator (PTI), and that an interface between two such PTIs can support edge states that are free from scattering. The conservation of the valley degree of freedom enables efficient in- and out-coupling of light between the free space and the photonic structure. The topological protection of the edge waves can be utilized for designing arrays of resonant time-delay photonic cavities that do not suffer from reflections and cross-talk.

*Introduction:*

The discovery of topological phases of light has been one of the most exciting developments in photonics [ 1, 2, 3, 4, 5, 6, 7, 8, 9, 10, 11] in the past decade. It followed the time-honored path of translating the concepts from condensed matter physics into the language of optical sciences, followed by developing novel applications based on those concepts. Photonic topological insulators (PTI) can be viewed as the extension of topological insulators [ 12, 13, 14, 15, 16, 17, 18, 19, 20] into the realm of optics. One potential application of PTIs is to utilize the reflections-free propagation of topologically protected edge waves (TPEWs) that exist either at the PTI's edge [ 3, 4, 5] or at the interface between two different PTIs [ 6, 7, 10] for developing robust optical delay lines for large-scale photonic integrations.

Specific implementations of PTIs vary considerably, and can utilize large coupled optical resonators [ 3, 4], wavelength-scale photonic structures [ 1, 2, 10, 21], or metacrystals [ 6]. To date, most of the wavelength and sub-wavelength scale PTI concepts utilized metals. For example, metallic metamaterials comprised of split-ring resonators [ 6] and meta-waveguides comprised of an array of metal rods attached to one of the two confining metal plates [ 10, 21] have been used to emulate the binary spin degrees of freedom (DOF) by ensuring that the two polarization states of light, the transverse electric (TE) and transverse magnetic (TM) modes, propagate with the same speed. This, property, known as spin-degeneracy [ 6, 7], is challenging to achieve without using metals. Avoiding metals is crucial if the spectral range of sub-wavelength PTI is going to be extended beyond the THz/mid-infrared portions of the electromagnetic spectrum.

In this Letter we demonstrate that an all-dielectric PTI can be developed by relying on a binary degree of freedom other than the spin. In designing the structure, we borrow the concept of the *valley* DOF from a rising field of valleytronics [ 22, 23, 24, 25, 26, 27]. It has been theoretically shown [ 22, 23, 24] that the *valley* DOF in any graphene-like material behaves as a spin-like binary DOF. Specifically, the angular rotation of the electron wavefunction in the $K$ or $K'$ valleys of the band structure generates an intrinsic magnetic moment [ 23] analogous to that produced by the electron spin. This similarity between the valley and spin DOF enables the quantum-valley Hall (QVH) effect [ 27] analogous to the quantum-spin Hall effect [ 12]. The latter effect manifests itself in the existence of the spin-locked (chiral) edge states at the graphene's edge which are immune to scattering by non-magnetic defects. Similarly, the QVH effect manifests itself in valley-locked chiral edge states that exist at the domain walls between, for example, AB and BA-stacked electrically biased bilayer graphene [ 25, 26, 27].

The analogy between the two effects and the utility of the valley DOF can be appreciated by comparing the mechanisms of topological protection in these two cases. In QSH effect, at the

Fermi level in the topological band gap, two edge states propagating in the opposite directions are locked to the spin DOF. As a consequence, a non-magnetic defect is unable to scatter a forward propagating edge state back (or vice versa) because no spin-flipping can take place during the scattering process. Similarly, for QVH effect, there exists a broad class of defects defined by their symmetry that act as 'non-magnetic' perturbations because they do not cause inter-valley scattering. Therefore, the edge states in the topological band gap are topological protected against backscattering in the presence of such defects.

Specifically, we show that the QVH effect can be emulated in an all-Si hexagonal photonic crystal with broken inversion symmetry (analogous to an AB-stacked electrically biased bilayer graphene) as shown in the inset of Fig.1(b). It thereby leads to the suppression of the inter-valley scattering and the topological protection in such photonic systems under a broad class of photonic lattice perturbations. The valley DOF utilized in QVH effect is much easier to realize in photonics than the spin DOF. For example, earlier work employed TE and TM polarizations to emulate the spin DOF [6, 7, 10]. A single TM polarization was employed for constructing the spin DOF in a recent all-dielectric design, but at the expense of using an enlarged (triple-sized) unit cell [28]. In contrast, as we show in this work, just a single (TE) polarization of the photonic modes is needed for constructing the valley DOF in a fairly simple photonic structure shown in Fig.1(b), where, by breaking the inversion symmetry of a unit cell, a controllable bandgap separating different topological phases of propagating light can be achieved.

We also demonstrate that an interface between two such quantum-valley-Hall (QVH) PTIs with different symmetry-breaking geometries [see Figs.2(a,b) for a typical example] supports highly-confined TPEWs. Topological protection against backscattering enables near-perfect out-coupling efficiency of TPEWs into vacuum as illustrated in Figs.3(a,b) despite their tight spatial confinement. Moreover, TPEWs can be used to confine light waves circulating inside an arbitrarily shaped/sized embedded defects exemplified by a quasi-random cavity shown in Fig.4(a). When placed in close proximity of a domain wall separating two different QVH-PTIs, such defect cavities can act as robust reflection-free optical delay lines.

*Design of a valley Hall all-dielectric photonic topological insulator:*

The starting point of the design is an unperturbed photonic graphene [29] comprised of a two-dimensional hexagonal lattice of circular Si rods with lattice constant $a_0$. The unit cell of the photonic graphene is shown in the inset of Fig.1(a). As in any uniaxial structure, the eigenmodes propagating in the $(x,y)$ plane can be classified as TE ($\boldsymbol{E}_\perp, H_z$) and TM ($\boldsymbol{H}_\perp, E_z$). Here we restrict the discussion to the TE modes; the representative field component can be thereby expanded with Bloch ansatz:

$$H_z(\boldsymbol{r}_\perp, t) = \sum_{n,\boldsymbol{k}_\perp} a^n(\boldsymbol{k}_\perp) h_z^{n,\boldsymbol{k}_\perp}(\boldsymbol{r}_\perp) e^{i\boldsymbol{k}_\perp \cdot \boldsymbol{r}_\perp - i\omega_n(\boldsymbol{k}_\perp)t} + c.c. \quad (1)$$

where the $n = 1,2$ index refers to lower (upper) photonic bands, and $h_z^{n,\boldsymbol{k}_\perp}$ is the normalized field profiles chosen to be periodic in the $\boldsymbol{r}_\perp = (x,y)$ plane. The normalized electric field components of the modes can be obtained from Eq.(1) from the following relation: $\boldsymbol{e}_\perp^{n,\boldsymbol{k}_\perp}(\boldsymbol{r}_\perp) = \frac{ic}{\omega_n \epsilon(\boldsymbol{r}_\perp)}[(ik_y + \partial_y)h_z^{n,\boldsymbol{k}_\perp}; -(ik_x + \partial_x)h_z^{n,\boldsymbol{k}_\perp}]$. The eigenfrequencies $\omega_n(\boldsymbol{k}_\perp)$ of the relevant two modes are calculated using COMSOL Multiphysics as shown in Fig.1(a) and highlighted in red, where $\boldsymbol{k}_\perp = (k_x, k_y)$ belongs to the first Brillouin zone (BZ). The boundaries and high symmetry points of the BZ are shown in the inset of Fig.1(a).

The presence of $C_{3v}$ wave vector symmetry group of the hexagonal lattice [30] results in cone-like dispersion curves in the non-equivalent $K$ and $K'$ "valleys" [22, 23, 24] of the BZ. The intersections of these cone-like dispersion curves at the $K$ and $K'$ points are known as the Dirac points; each of them is doubly-degenerate. This degeneracy at the Dirac frequency $\omega_D$ allows us to choose the orthonormal orbital basis of the right- or left-hand circular polarizations (RCP and LCP). The field profiles of the RCP and LCP modes for the unperturbed photonic graphene are shown in Fig.1(c), and they are invariant under a $2\pi/3$ rotation, $\mathcal{R}_3$, along z-axis (See Supplementary Information for more details).

In order to investigate the topological aspect of the modes, an effective Hamiltonian description must be analytically derived, and the connection to the commonly known Hamiltonian of 2D topological materials must be established. This effective Hamiltonian $\mathcal{H}(\boldsymbol{k}_\perp)$, which is a function of the in-plane wavenumber, will be expressed in the basis of the RCP/LCP states in close proximity of the $K$ and $K'$ points of the BZ. Of course, such effective Hamiltonian is an approximation because it utilizes a small number of propagating eigenmodes of the photonic structure, and because its validity is justified only near the high symmetry points of the BZ. However, it serves two purposes: (a) qualitative explanation of the emergence of the bandgap due to the inversion symmetry breaking, and (b) calculation of the relevant topological invariants, known as the valley-Chern indices [31] in condensed matter physics. The existence of such non-vanishing topological indices are crucial for the existence of the TPEWs at an interface between two QVH-PTIs. The analytic calculation of $\mathcal{H}(\boldsymbol{k}_\perp)$ in the presence of the inversion symmetry breaking deformation of a Si rod [exemplified by an inset in Fig.1(b)] will be based on the first-order perturbation theory [32].

We start the analytic calculation by defining the corresponding expansion basis and the state vectors of the effective Hamiltonian. If the amplitudes of the RCP/LCP modes in the corresponding valleys are $a_{K(K')}^R$ and $a_{K(K')}^L$, respectively, then any eigenstate can be expressed as a state vector $\mathbf{U}_{K(K')} = [a_{K(K')}^R; a_{K(K')}^L]$. The degenerate orbital expansion basis $\mathbf{U}_R = [1; 0]$ and $\mathbf{U}_L = [0; 1]$ can be defined according to the symmetry property (invariance under $\mathcal{R}_3$) of the RCP and LCP fields which is mathematically expressed as $\mathcal{R}_3 \mathbf{U}_{R,L} = \exp(\mp 2\pi i/3) \mathbf{U}_{R,L}$. An extended state vector combining both valleys can be defined as $\boldsymbol{\Psi} = [\mathbf{U}_K; \mathbf{T}\mathbf{U}_{K'}]$, where the transformation matrix $\mathbf{T}$ swaps the RCP and LCP orbital states to follow the commonly used convention that pairs up the amplitudes of the modes ($a_K^{R(L)}$ and $a_{K'}^{L(R)}$) connected through the time-reversal transformation. The unperturbed Hamiltonian $\mathcal{H}(\boldsymbol{k}_\perp) \equiv \mathcal{H}_0(\delta \boldsymbol{k})$ near the Dirac points in this basis can then be expressed [12] as

$$\mathcal{H}_0(\delta \boldsymbol{k}) = v_D \left( \delta k_x \hat{\tau}_z \hat{\sigma}_x + \delta k_y \hat{\tau}_0 \hat{\sigma}_y \right), \tag{2}$$

where $v_D$ is the group velocity, and $\delta \boldsymbol{k} = (\delta k_x, \delta k_y) \equiv \boldsymbol{k}_\perp - \boldsymbol{k}_D$ is the distance from the Dirac points defined as $\boldsymbol{k}_D = \pm \boldsymbol{e}_x 4\pi/3 a_0$ for the $K$ and $K'$ points, respectively. The Pauli matrices $\hat{\sigma}_{x,y,z}$ and $\hat{\tau}_{x,y,z}$ act on the orbital and valley state vector, respectively, $\hat{\sigma}_0, \hat{\tau}_0$ are the corresponding identity matrices, and $\hat{\tau}_i \hat{\sigma}_i \equiv \hat{\tau}_i \otimes \hat{\sigma}_i$ is the shorthand for the Kronecker product.

Next we introduce the perturbation of a unit cell accomplished by the deformation of the Si rods. It has been shown [21] that any geometric perturbation that breaks the inversion ($P$)-symmetry yet preserves the $C_3$ point symmetry does not couple the RCP and LCP modes. It means that the perturbation matrix (responsible for lifting the degeneracy of the RCP and LCP orbital states) is diagonalized:

$$\mathcal{H}_P = \omega_D \Delta_P \hat{\tau}_0 \hat{\sigma}_z \tag{3}$$

The proportionality of $\hat{\tau}_0$ simply shows that the effect of the perturbation is the same for both valleys, and it is a consequence of the time-reversibility. An example of such a geometric perturbation is shown in Fig.1(b). The corresponding field profiles after the perturbation are shown in Fig.1(d). Together with Eqs.(2,3), one can obtain the photonic band structure of the perturbed system by calculating the eigenfrequency, $\Omega(\delta k) \equiv \omega(\delta k) - \omega_D$, of the matrix equation $\mathcal{H}(\delta k)\Psi = \Omega(\delta k)\Psi$, where $\mathcal{H} = \mathcal{H}_0 + \mathcal{H}_P$. The size of the bandgap $\Delta\omega = 2\omega_D|\Delta_P|$ is proportional to the perturbation strength $\Delta_P$ which can be determined from the first-order perturbation theory [32]:

$$2\Delta_P \equiv \Delta^{RR}_{P,K(K')} - \Delta^{LL}_{P,K(K')} = -\int_V \Delta\epsilon(\mathbf{r}_\perp) \cdot (|\mathbf{e}^R_\perp|^2 - |\mathbf{e}^L_\perp|^2)\, dV, \quad (4)$$

where $\Delta V$ is the pertubated volume, and $\Delta\epsilon(\mathbf{r}_\perp) = \pm(\epsilon_{Si} - 1)$ is the changing permittivity after perturbation (circular to triangular rod). The $\pm$ sign of $\Delta\epsilon(\mathbf{r}_\perp)$ depends on whether the vacuum region is replaced by Si or vice versa. It can be observed from Fig.1(c) that the sign of $\Delta_P$ depends on the orientation of the triangular rod. For example, the triangular rod with one of the three vertices pointing toward $+y$ as shown in Fig.1(b) has the most negative $\Delta_P$ (indicating that $\Delta^{RR}_{P,K(K')} < \Delta^{LL}_{P,K(K')}$), whereas a triangular rod rotated by 180 degrees has the largest positive $\Delta_P$ ($\Delta^{RR}_{P,K(K')} > \Delta^{LL}_{P,K(K')}$).

Although the band structures and the eigenfrequencies $\Omega(\delta k)$ of the perturbed system with the opposite signs of $\Delta_P$ are identical to each other, the topological indices of the propagating modes in these two photonic structures are not. The nontrivial topological property of the modes can be characterized by the nonvanishing valley-Chern indices [31], $C^{(v)} \neq 0$. By definition [33], $C^{(v)} = \int_{BZ(v)} d^2\delta k\, [\nabla_{\delta k} \times \mathbf{A}(\delta k)]_z / 2\pi$, where $v = K, K'$ is the valley label, and $BZ(v)$ is half of the BZ corresponding to $k_x > 0 (< 0)$ for $v = K(K')$, respectively. The local Berry connection [34, 35, 36] is calculated as $\mathbf{A}(\delta k) = -i\psi^\dagger_v(\delta k) \cdot \nabla_k \psi_v(\delta k)$, where $\psi_K = \mathbf{U}_K$ and $\psi_{K'} = \mathbf{T}\mathbf{U}_{K'}$ are projections onto the $v$ valley subspace of the full spinor $\Psi(\delta k)$ as the eigenvector with the corresponding eigenfrequency below the bandgap. Using the effective Hamiltonian of Eqs.(2,3), we calculated the non-vanishing valley-Chern indices to depend on the specific valley and on the sign of $\Delta_P$ according to $2C^{(K,K')} = \pm 1 \times \text{sgn}(\Delta_P)$.

The ability to control the sign of $\Delta_P$ by simply rotating the triangular rods of a QVH-PTI allows one to create a topological cladding that supports topologically protected edge waves (TPEWs) [21] at the interface between two QVH-PTIs with opposite signs of $\Delta_P$. According to the *bulk-boundary* correspondence [37], the difference between the number of forward-moving modes and the number of backward-moving modes equals the difference of the valley-Chern indices. For example, an interface along the zigzag direction shown in Fig.2(a) has $\Delta_P > 0 (< 0)$ in the upper(lower) domain; consequently, there is a forward-moving TPEW at the $K$ point because $\Delta C^{(K)} = C^{(K)}_{\text{upper}} - C^{(K)}_{\text{lower}} = 1$, and a backward-moving TPEW at the $K'$ point because $\Delta C^{(K)} = -1$. The dispersions of TPEW for such interfacing are calculated using COMSOL Multiphysics and shown in Fig.2(c). According to the propagation direction of TPEWs, here we refer to the interfaces shown in Figs.2(a,b) as 'positive-type' and 'negative-type', respectively. The TPEWs are strongly localized near the interface between PTIs with the opposite signs of $\Delta_P$ on either side of the interface.

The most remarkable property of the TPEWs that can be observed from Fig.2(c) is that, for a given valley, there is only one edge mode propagating in the direction that is "locked" to the valley. Therefore, if no inter-valley scattering takes place, then a forward-propagating edge mode cannot back-scatter. This property of the edge modes is responsible for their topological protection. Below we discuss the physical origin of the suppression of inter-valley scattering along with two applications TPEWs can offer: (i) to reflections-free out-coupling into vacuum, and (ii) to designing topologically protected random cavities that can be utilized for time-delaying optical pulses.

*Application: valley conservation at a zigzag termination of a PTI waveguide*

Of course, TPEWs are not unique in their localization: a variety of guided modes can propagate in channels separating any two photonic crystals that have a bandgap around the desired wavelength. The advantage of using topological claddings which, in addition to having a bandgap, also possess different signs of the valley-Chern index, is that the resulting TPEWs cannot back-scatter as long as they stay within the same valley (i.e. there is no inter-valley scattering). Such topological protection presents an interesting opportunity for reflection-free out-coupling of the TPEWs into the vacuum region. The key to such efficient out-coupling is the absence of inter-valley scattering at the specific termination of the photonic structure. To anticipate the types of terminations that result in a vanishing inter-valley scattering, one needs to calculate the field overlap between two valleys' TPEWs. It turns out that the inter-valley field overlap vanishes under the so-called zigzag-type perturbations, but remains nonzero when the perturbation is armchair-type (See SI for detailed discussions). It is thereby expected that a TPEW cannot scatter into a different valley when encountering a termination which is along the zigzag [38] direction as shown in Figs.3(a,b).

We used COMSOL simulations to demonstrate the perfect matching of TPEWs into vacuum. The setup shown in Figs.3(a,b) consists of three in-line phase-matched point dipoles and used to launch forward-propagating TPEWs along the positive- and negative-type interfaces, respectively. We observe from Figs.3(a,b) that the TPEWs are smoothly out-coupled into vacuum without any reflections. Notably, the direction of the outgoing beam into vacuum depends on the effective refractive indices of the QVH-PTI waveguides: the light launched into vacuum from a positive-type interface between two PTIs refracts according to the standard (positive-index) Snell's Law as shown in Fig.3(a). On the other hand, the negative-type interface launches negatively-refracted light into vacuum as shown in Fig.3(b). However, in both cases reflections are absent. Note that the time-reversal symmetry guarantees that perfect coupling into TPEWs from vacuum can be obtained by properly designing an optical beam impinging from the vacuum-side with its profile accurately matched to Figs.3(a,b).

To verify that perfect out-coupling of TPEWs is indeed related to the conservation of the valley DOF, we have constructed an alternative termination of the QVH-PTI waveguide that does not suppress inter-valley scattering. Such armchair terminations are shown in Figs.3(c,d) as red vertical lines. The reflection calculated using COMSOL simulations shown in Figs.3(c,d) are $R = 0.16$ for the positive-index TPEWs and $R = 0.99$ for the negative-index TPEWs. This example demonstrates that, in the absence of complete suppression of inter-valley scattering afforded by a zigzag interface, very strong reflection should be expected at an impedance-mismatched interfaces such as between a photonic crystal and vacuum.

It is worth to point out that although a TPEW is immune to the reflection from a sharp bend [10, 21] inside a PTI waveguide, the reflection-free feature is by no means guaranteed when a TPEW hits the edge of a PTI waveguide. Poor coupling between guided and free-space modes is a generic problem in the field of optical communications which can be generally formulated in terms of impedance mismatch between the waveguide and vacuum. This problem, however, can be overcome in a QVH-PTI waveguide by terminating the sample along a zigzag direction as shown in Figs.3(a,b). Even though the vacuum does not have an appropriate valley DOF either, the inter-valley scattering does not take place because of the symmetry properties of a zigzag-type perturbation which results in a vanishing field overlap between the two TPEWs belonging to different ($K$ and $K'$) valleys.

*Application: robust delay lines based on reflection-free random cavities*

When two PTIs with the opposite signs of $\Delta_P$ are embedded in each other, as shown in Fig.4(a), multiple zigzag interfaces between the two phases emerge. Those interfaces can guide edge waves along multiple paths without back-scattering. Therefore, such a mixture of $\Delta_P > 0$ and $\Delta_P < 0$ phases becomes a resonant cavity when imbedded inside a PTI. The number of TPEW modes supported by a cavity depends its size, as well as on the relative abundance of the two phases. When such two-phase cavities are placed in close proximity of a bus waveguide comprised of a straight interface between two PTIs as shown in Fig.2(b), the trapped TPEWs can weakly couple to the passing edge mode. Because of the suppression of inter-valley scattering, the two-phase cavity is directionally coupled to the bus waveguide. Therefore, one can envision a robust reflections-free delay line that relies on directional coupling for suppressing reflections. Unlike standard directional couplers that must be multiple wavelengths long to avoid back-scattering, the proposed topologically protected delay line can be rather compact.

The valley conservation or the suppression of the inter-valley scattering can be understood by calculating the field overlaps between the modes of two valleys. It has been shown [21] that within the perturbed volume of a single unit cell, the field overlaps of the modes belonging to different valleys vary between positive and negative values; advantageously, such variations tend to average out (See SI for details). Besides, the field overlap integrals in the neighboring cells also tends to cancel each other out. In fact, as demonstrated in SI, a perturbation consisting of rotating three triangular rods in a row along any zigzag directions (zigzag-type perturbation) in the exact same way results in a perfect cancellation of the individual perturbations. Therefore, a perturbation with random rotations among cells results only a higher-order coupling: $O[\Delta_P^2]$ (coming from non-zigzag-type perturbations) between two valley modes as opposed to the first-order coupling caused by $C_3$ perturbation [Fig.1(b)] responsible for a topological bandgap: $\Delta\omega = 2\omega_D|\Delta_P|$. From this discussion, one can expect that the coupling strength between two valleys' modes from a random perturbation are considerably smaller than $\Delta_P$.

To confirm this conjecture, we perform an eigenfrequency simulation of a random cavity with its geometry shown in Fig.4(a). Because of the unbroken time-reversal symmetry and the fact that two valleys' modes are one-to-one time-reversal partners, one expects that the discrete eigenmodes of such cavity with localized fields must all be doubly-degenerate if the inter-valley interaction is strictly zero. That implies that all eigenvalues must come as pairs, i.e. $\omega_+^{(i)} = \omega_-^{(i)} \equiv \omega^{(i)}$, where the superscript labels the pair. If any finite frequency splitting $\Delta\omega^{(i)} \equiv \omega_+^{(i)} - \omega_-^{(i)}$ emerges, then it must be attributed to inter-valley scattering and reflect the coupling strength between the two valleys' modes. For a specific two-phase cavity shown in Fig.4(a), we have calculated using COMSOL eigenvalue modeling that the resultant normalized eigenfrequencies are $\omega_\pm^{(1)} a_0/2\pi c = 0.42836(1 \pm 0.016\%)$ and $\omega_\pm^{(2)} a_0/2\pi c = 0.46831(1 \pm 0.019\%)$. Note the near-perfect degeneracy of the modes in each of the pairs: the normalized frequency splitting in each of the mode pairs is very small compared with the topological bandgap, i.e. $\Delta\omega^{(1)}/2\omega_D\Delta_P = 0.001$ and $\Delta\omega^{(2)}/2\omega_D\Delta_P = 0.0013$, respectively. Therefore, a typical isolated two-phase cavity shown in Fig.4(a) is indeed valley-preserving, and can be potentially used as a whispering-gallery-type resonator for a reflection-free optical delay line. As discussed earlier, when such a cavity is coupled to a propagating TPEW, it acts as a directionally coupled resonator. The transmission of a TPEW propagating past, and coupled to, such a resonator is spectrally flat near the resonance frequencies $\omega^{(i)}$ of the resonator. However, its phase undergoes rapid changes at $\omega^{(i)}$'s because the propagating TPEW couples to the trapped TPEWs inside the two-phase cavity and, effectively, spends more time circulating inside the directionally coupled resonator.

An example of a two-phase topological cavity (emulating a directionally coupled resonator) coupled to a TPEW-supporting straight interface between two PTIs (emulating a bus waveguide) is shown in Fig.4(b), where the cavity is separated by two rows from the straight interface between PTIs. If the coupling rate $\kappa$ between the bus waveguide and the directionally coupled resonator greatly exceeds the inter-valley coupling rate $\Delta\omega^{(i)}$ inside the cavity, and this external coupling does not significantly increase $\Delta\omega^{(i)}$, then the above mentioned unity transmission $T(\omega) \approx 1$ and the rapid change of the phase $\phi(\omega)$ of the transmitted signal with the complex-valued coefficient $t = \sqrt{T}\exp(i\phi)$ are both anticipated. The results of a COMSOL simulation presented in Fig.4(c) confirm these conjectures. The transmission is mostly near unity and above $T = 0.9$ across the bandgap frequency range. The delay time $\tau_{\text{delay}} = d\phi/d\omega - d\phi_0/d\omega$, where $\phi(\omega)$ and $\phi_0(\omega)$ are calculated with and without the resonant cavity, is normalized to $a_0/c$ and plotted in Fig.4(c) for the entire bandgap frequency range. The peak of time delay as large as $\tau_{\text{delay}} \approx 600a_0/c$, from a two-phase cavity that is only $8 \times 4$ rows in size, without compromising high transmission, are predicted by our simulation.

The zoom-in of $\tau_{\text{delay}}$ in the spectral vicinity of $\omega^{(1)}$ plotted in the inset of Fig.4(c) is used to estimate $\kappa$. From the linewidth of the resonance, the coupling rate between the waveguide and the cavity is estimated as $\kappa \approx 17\Delta\omega^{(1)}$, thus validating the above $\kappa \gg \Delta\omega^{(1)}$ assumption. The effective optical length of the cavity at the resonance frequency $\omega^{(1)}$ can also be estimated from as $l_{\text{eff}} \equiv v_D\tau_{\text{delay}} \approx 208a_0$. Finally, high transmission $T > 90\%$ for all frequencies across the resonance indicates that the inter-valley scattering indeed remains weak, thus preventing reflections even for a high-Q ($Q \approx 416$) time-delay line.

While it is too early to predict the full potential of such topologically-protected random cavities, below we offer some arguments that may set such cavities apart from the standard directionally coupled resonators used, for example, in constructing ultra-compact optical buffers and delay lines [39]. One key property of a QVH-PTI random cavity is that it confines the circulating TPEWs inside its entire area $A$. Therefore, the number of modes in the cavity scales as $N \sim A$. The implication of this scaling is that the mode density of a random cavity in a given area is larger than a traditional Si-ring whispering-gallery resonator. In the latter, the modes' number scaling is $N \sim \sqrt{A}$ because light is guided along the ring's circumference while leaving the inner area unutilized. This unfavorable scaling leads to the increase of the footprint of the device when one attempts to increase the time-delay bandwidth by using a series of ring resonators with different diameters.

To understand how a random cavity can alleviate this challenge, one might envision using a spaced sequence of random cavities of the same size but different from each other in the way the "flipped" (i.e. $\Delta_P < 0$) triangular rods are embedded in the $\Delta_P > 0$ photonic matrix. Inside each cavity, one half of the rods can be randomly flipped to support a certain number of cavity eigenmodes. The neighboring cavity would then have a different combination of the flipped rods, and support the same number of modes but with different eigenfrequencies. By using multiple combination of the flipped rods and stacking such random cavities in a sequence, the entire bandgap region would be spanned by the isolated time-delay peaks shown in Fig.4(c), thereby broadening the operational bandwidth.

In conclusion, we have designed a new photonic topological insulator analogous to the quantum-valley-Hall effect using all-dielectric photonic crystal. The principle of valley conservation enables QVH-PTIs to be used for reflection-free guiding along arbitrary paths and for robust optical delays with random cavities. The fact that only a single TE polarization is required for constructing the valley degrees of freedom greatly releases the previous design restriction demanding to use both TE and TM polarization [ 6, 7, 10, 21]. The underlying physics responsible for the valley conservation under zigzag-type perturbation gives the opportunity for highly efficient external excitation of topologically-protected edge waves from the vacuum to a QVH-PTI waveguide. For these reason, a QVH-PTI can potentially serve as an important component in optical communication and integrated Si-photonics.

## References
1. Wang, Z., Chong, Y., Joannopoulos, J. & Soljačić, M., Reflection-free one-way edge modes in a gyromagnetic photonic crystal. *Phys. Rev. Lett.* **100**, 13905 (2008).
2. Wang, Z., Chong, Y., Joannopoulos, J. D. & Soljačić, M., Observation of unidirectional backscattering-immune topological electromagnetic states. *Nature* **461**, 772-775 (2009).
3. Hafezi, M., Demler, E. A., Lukin, M. D. & Taylor, J. M., Robust optical delay lines with topological protection. *Nature Physics* **7** (11), 907-912 (2011).
4. Hafezi, M., Mittal, S., Fan, J., Migdall, A. & Taylor, J. M., Imaging topological edge states in silicon photonics. *Nature Photonics* **7**, 1001–1005 (2013).
5. Rechtsman, M. C. *et al.*, Photonic Floquet topological insulators. *Nature* **496** (7444), 196-200 (2013).
6. Khanikaev, A. B. *et al.*, Photonic topological insulators. *Nature Materials* **12** (3), 233-239 (2013).
7. Chen, W.-J. *et al.*, Experimental realization of photonic topological insulator in a uniaxial metacrystal waveguide. *Nat Commun* **5**, 5782 (2014).
8. Gao, W. *et al.*, Topological Photonic Phase in Chiral Hyperbolic Metamaterials. *Phys. Rev. Lett.* **114**, 037402 (2015).
9. Liu, F. & Li, J., Gauge Field Optics with Anisotropic Media. *Phys. Rev. Lett.* **114**, 103902 (2015).
10. Ma, T., Khanikaev, A. B., Mousavi, S. H. & Shvets, G., Guiding Electromagnetic Waves around Sharp Corners: Topologically Protected Photonic Transport in Metawaveguides. *Phys. Rev. Lett.* **114** (12), 127401 (2015).
11. Lu, L., Joannopoulos, J. & Soljačić, M., Topological photonics. *Nature Photonics* **8**, 821 (2014).
12. Kane, C. L. & Mele, E. J., Quantum Spin Hall Effect in Graphene. *Phys. Rev. Lett.* **95**, 226801 (2005).

13. König, M. *et al.*, Quantum spin Hall insulator state in HgTe quantum wells. *Science* **318** (5851), 766-770 (2007).
14. Bernevig, B. A. & Zhang, S.-C., Quantum Spin Hall Effect. *Phys. Rev. Lett.* **96**, 106802 (2006).
15. Fu, L., Kane, C. L. & Mele, E. J., Topological Insulators in Three Dimensions. *Phys. Rev. Lett.* **98**, 106803 (2007).
16. Hsieh, D. *et al.*, A topological Dirac insulator in a quantum spin Hall phase. *Nature* **452** (7190), 970-974 (2008).
17. Roth, A. *et al.*, Nonlocal transport in the quantum spin Hall state. *Science* **325** (5938), 294-297 (2009).
18. Zhang, H. *et al.*, Topological insulators in Bi2Se3, Bi2Te3 and Sb2Te3 with a single Dirac cone on the surface. *Nature Physics* **5** (6), 438-442 (2009).
19. Chen, Y. *et al.*, Experimental realization of a three-dimensional topological insulator, Bi2Te3. *Science* **325** (5937), 178-181 (2009).
20. Xia, Y. *et al.*, Observation of a large-gap topological-insulator class with a single Dirac cone on the surface. *Nature Physics* **5** (6), 398-402 (2009).
21. Ma, T. & Shvets, G., Scattering-Free Optical Edge States between Heterogeneous Photonic Topological Insulators. *arXiv preprint arXiv:1507.05256* (2015).
22. Rycerz, A., Tworzydło, J. & Beenakker, C., Valley filter and valley valve in graphene. *Nature Physics* **3** (3), 172-175 (2007).
23. Xiao, D., Yao, W. & Niu, Q., Valley-Contrasting Physics in Graphene: Magnetic Moment and Topological Transport. *Phys. Rev. Lett.* **99**, 236809 (2007).
24. Yao, W., Xiao, D. & Niu, Q., Valley-dependent optoelectronics from inversion symmetry breaking. *Phys. Rev. B* **77**, 235406 (2008).
25. Zhang, Y. *et al.*, Direct observation of a widely tunable bandgap in bilayer graphene. *Nature* **459**, 820 (2009).
26. Ju, L. *et al.*, Topological valley transport at bilayer graphene domain walls. *Nature* **520**, 650 (2015).
27. Kim, Y., Choi, K., Ihm, J. & Jin, H., Topological domain walls and quantum valley Hall effects in silicene. *Phys. Rev. B* **89** (8), 085429 (2014).
28. Wu, L.-H. & Hu, X., Scheme for Achieving a Topological Photonic Crystal by Using Dielectric Material. *Phys. Rev. Lett.* **114**, 223901 (2015).
29. Szameit, A., Rechtsman, M. C., Bahat-Treidel, O. & Segev, M., PT-symmetry in honeycomb photonic lattices. *Phys. Rev. A* **84**, 021806(R) (2011).
30. Malterre, D. *et al.*, Symmetry breaking and gap opening in two-dimensional hexagonal lattices. *New Journal of Physics* **13** (1), 013026 (2011).
31. Ezawa, M., Topological Kirchhoff law and bulk-edge correspondence for valley Chern and spin-valley Chern numbers. *Phys. Rev. B* **88**, 161406 (2013).
32. Pozar, D. M., *Microwave Engineering*, 4th ed. (Wiley Global Education, 2011).
33. Yang, Y. *et al.*, Time-Reversal-Symmetry-Broken Quantum Spin Hall Effect. *Phys. Rev. Lett.* **107**, 066602 (2011).

34. Thouless, D., Kohmoto, M., Nightingale, M. & Den Nijs, M., Quantized Hall conductance in a two-dimensional periodic potential. *Physical Review Letters* **49** (6), 405 (1982).
35. Simon, B., Holonomy, the quantum adiabatic theorem, and Berry's phase. *Physical Review Letters* **51** (24), 2167 (1983).
36. Sheng, D., Weng, Z., Sheng, L. & Haldane, F., Quantum spin-Hall effect and topologically invariant Chern numbers. *Physical review letters* **97** (3), 036808 (2006).
37. Mong, R. S. & Shivamoggi, V., Edge states and the bulk-boundary correspondence in Dirac Hamiltonians. *Physical Review B* **83** (12), 125109 (2011).
38. Dresselhaus, M. S., Dresselhaus, G. & Eklund, P. C., *Science of fullerenes and carbon nanotubes: their properties and applications* (Academic press, 1996).
39. Xia, F., Secaric, L. & Vlasov, Y., Ultracompact optical buffers on a silicon chip. *Nature Photonics* **1**, 65 (2007).

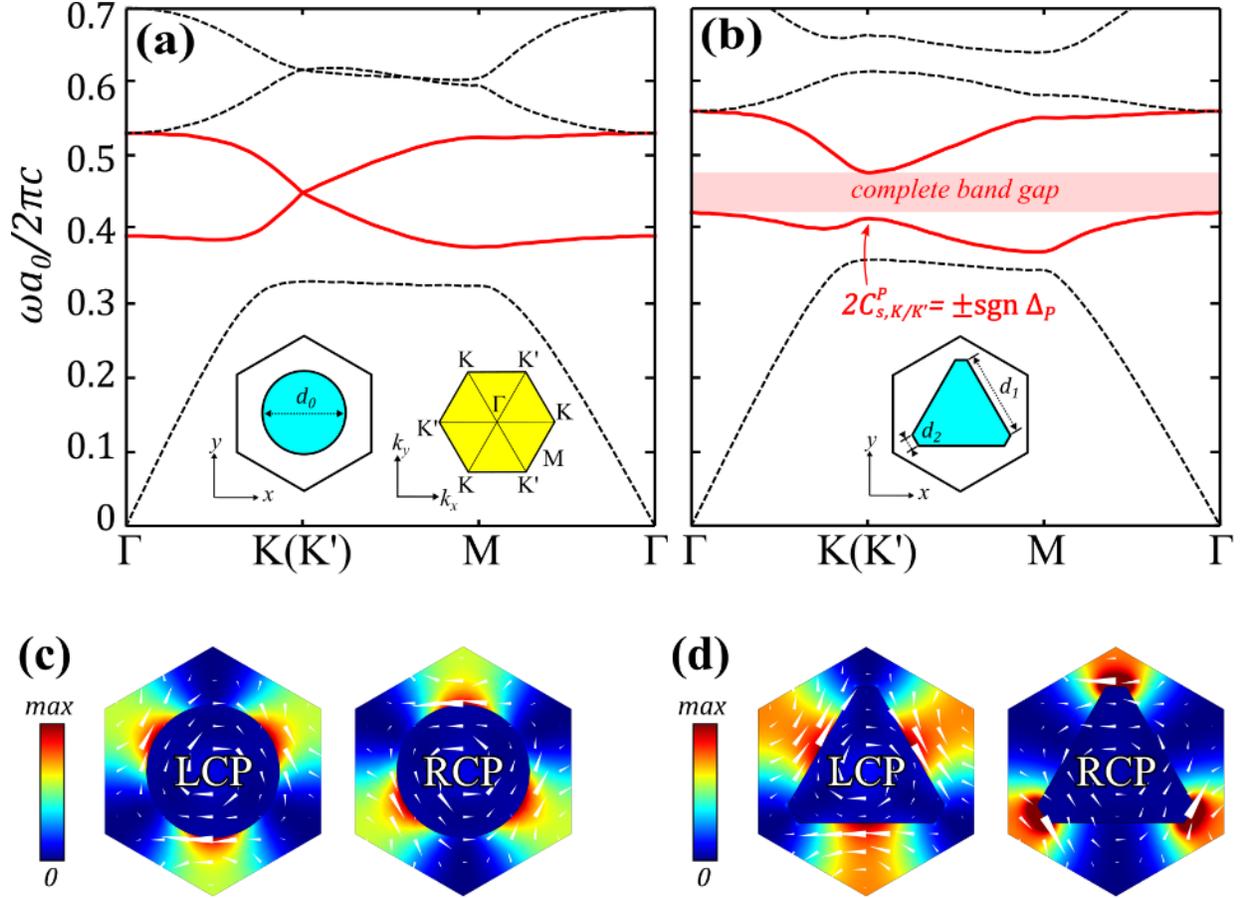

**Figure 1: Photonic band structure and electromagnetic modes of unperturbed and perturbed photonic structures.** (a) The photonic band structure of an unperturbed all-Si "photonic graphene". Red lines: the relevant photonic bands forming a Dirac points at the $K(K')$ point. Insets: the unit cell with a round Si rod in vacuum (left) and the first Brillouin zone (right). (b) The photonic band structure of a perturbed Si-PhC as a QVH-PTI. Inset: triangular Si rod. (c) Unperturbed and (d) perturbed field profiles of the LCP and RCP states at the $K$ point of a photonic graphene. Color: $|\mathbf{E}|^2$; Arrows: power flux. Parameters of the QVH-PTI: $\epsilon_{Si} = 13$, $d_0 = 0.615 a_0$, $d_1 = 1.65 a_0$, $d_2 = 0.11 a_0$, where $a_0$ is the lattice constant.

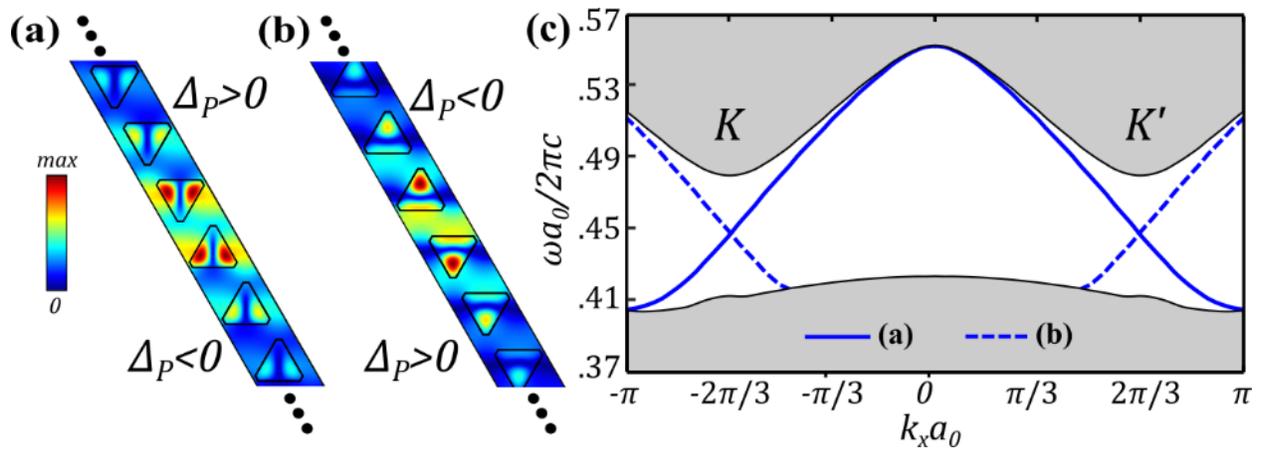

**Figure 2: Topologically protected edge waves (TPEWs) along zigzag interfaces between two QVH-PTI claddings.** **(a)** The super-cell used for COMSOL simulations: single cell along the propagation $x$-direction and 20 cells on each side of the interface. The field profile of the TPEW at the $K(K')$ point for the positive-type interface with group velocity being positive(negative). **(b)** Same as (a) but with the negative-type interface in which the triangular rods are flipped by 180 degrees. Color: $|H_z|$. **(c)** The photonic band structure of the structures in (a) and (b). Gray-shaded region: bulk modes; solid/dashed blue lines: TPEWs in (a)/(b).

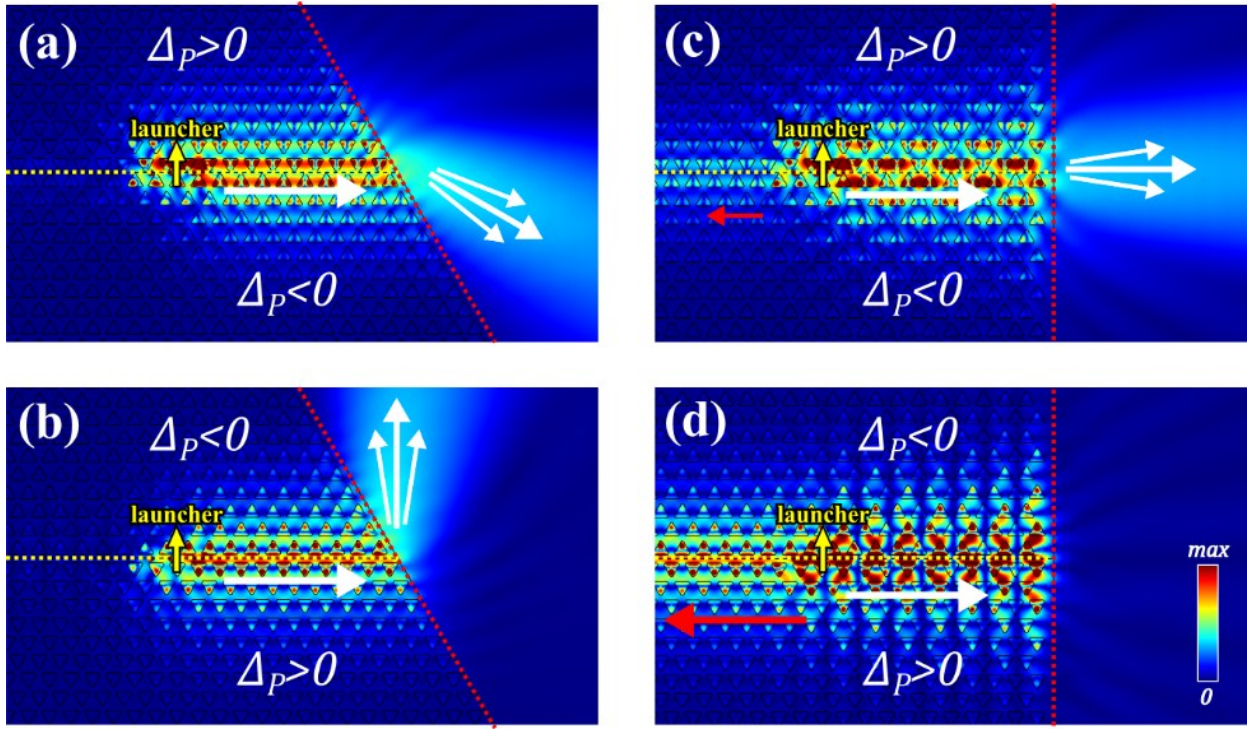

**Figure 3: Out-coupling of TPEWs into vacuum with different terminations. (a,b)** Zigzag terminations of the QVH-PTI waveguide. (a,b) Efficient out-coupling for zigzag terminations of the structures shown in Figs.2(a) and (b). **(c,d)** Same as (a,b) but with armchair terminations. White arrows: transmitted waves, red arrows: reflected waves. Color: $|H_z|$. Yellow- and red-dotted lines highlight the guiding interface and the termination, respectively.

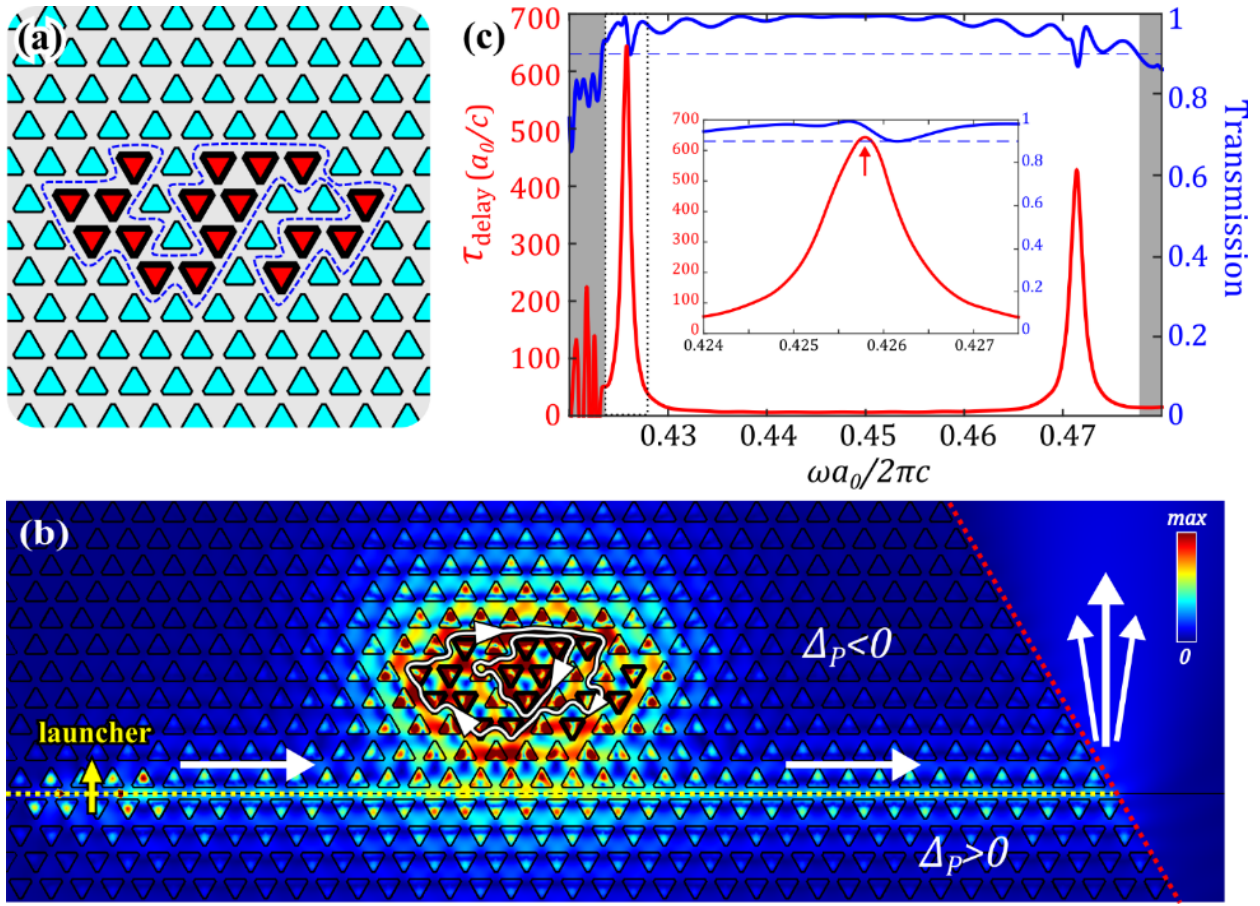

**Figure 4: Robust delay lines using whispering-gallery-type resonators with an arbitrary shape. (a)** Schematic of the isolated random two-phase cavity (red triangles: $\Delta_P > 0$ inside $\Delta_P < 0$ matrix). The blue-dashed line outlines the interface between the two phases. **(b)** Cavity in (a) is coupled to a QVH-PTI bus waveguide. White arrows: direction of the power flow. Color: $|H_z|$. Yellow- and red-dotted lines: guiding interface and the termination into vacuum. **(c)** Delay time $\tau_{\text{delay}}(\omega)$ and transmission $T(\omega)$. The gray-shaded region are outside the bandgap of QVH-PTIs; the smaller middle spectra are the zoom-in of the dashed-line box. The blue-dashed line marks the 90% transmission to guide eyes. The red arrow indicates the frequency where the field profiles of (b) is plotted. Parameters: same as in Fig.1(b).

# Supplementary Information

*RCP and LCP modes & the diagonalized perturbation Hamiltonian:*
In this section we illustrate the field distribution of RCP and LCP modes, and how their symmetry properties result in a zero coupling between them so that the perturbation Hamiltonian is simply diagonalized.

The origin of degenerate RCP and LCP of an unperturbed photonic graphene can be understood from studying the lowest bands of a vacuum hexagonal lattice as shown in Fig.S1(a). The corresponding reciprocal lattice in $k$-space is shown in Fig.S1(b). At the $K$ point, three adjacent cone-like bands (simple dispersion for vacuum) intersect. The $K$ point is where one constructs RCP and LCP modes which are made of the linear superposition of the three plane waves: $H_z^{(K)} = a_1 e^{i\boldsymbol{k}_1 \cdot \boldsymbol{r}} + a_2 e^{i\boldsymbol{k}_2 \cdot \boldsymbol{r}} + a_3 e^{i\boldsymbol{k}_3 \cdot \boldsymbol{r}}$, where $\boldsymbol{k}_i = \frac{4\pi}{3a_0}\left[\cos\frac{2\pi(i-1)}{3}\hat{\boldsymbol{x}} + \sin\frac{2\pi(i-1)}{3}\hat{\boldsymbol{y}}\right]$ [also indicated in Fig.S1(b)]. Next, a round Si rod like Fig.1(a) shows in the main texts is added into vacuum unitcell, and consequently lift the three-fold degeneracy of these three plane waves at the $K$ point. It is not difficult to see that a particularly symmetric orthonormal basis will diagonalized this perturbation:

$$H_z^{(K,\text{singlet})} = 1 e^{i\boldsymbol{k}_1 \cdot \boldsymbol{r}} + 1 e^{i\boldsymbol{k}_2 \cdot \boldsymbol{r}} + 1 e^{i\boldsymbol{k}_3 \cdot \boldsymbol{r}} \tag{S1.1}$$

$$H_z^{(K,\text{LCP})} = 1 e^{i\boldsymbol{k}_1 \cdot \boldsymbol{r}} + \eta e^{i\boldsymbol{k}_2 \cdot \boldsymbol{r}} + \eta^* e^{i\boldsymbol{k}_3 \cdot \boldsymbol{r}} \tag{S1.2}$$

$$H_z^{(K,\text{RCP})} = 1 e^{i\boldsymbol{k}_1 \cdot \boldsymbol{r}} + \eta^* e^{i\boldsymbol{k}_2 \cdot \boldsymbol{r}} + \eta e^{i\boldsymbol{k}_3 \cdot \boldsymbol{r}} \tag{S1.3}$$

where $\eta = \exp\left(i\frac{2\pi}{3}\right)$.

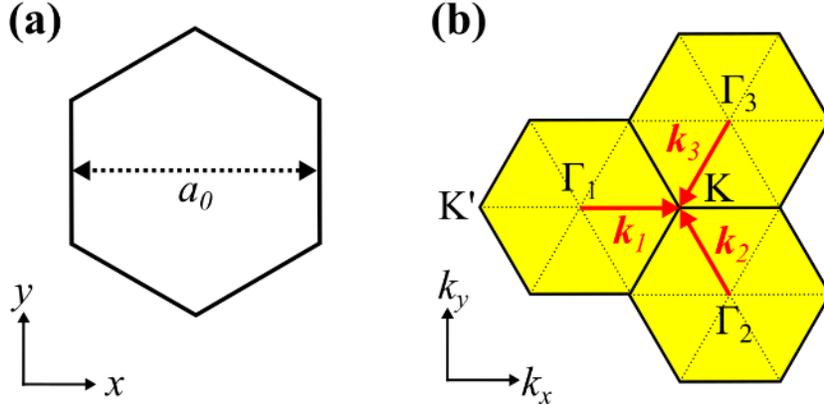

**Figure S1: Three-fold degeneracy of a hexagonal vacuum lattice at the $K$ point.** (a) A hexagonal vacuum lattice. (b) The corresponding reciprocal lattice.

As shown in Fig.S2, $H_z^{(K,\text{singlet})}$ has an ordinary standing-wave field distribution with the intensity mostly concentrated in the middle of a unitcell so that its eigenfrequency goes lower after the round-Si-rod perturbation. $H_z^{(K,\text{RCP})}$ and $H_z^{(K,\text{LCP})}$ on the other hand have a nod in intensity in the middle of a unitcell and thereby have higher eigenfrequency. Also note that the field profile of $H_z^{(K,\text{RCP})}$ is the mirror image of $H_z^{(K,\text{LCP})}$ and since the round-Si rod does not break the inversion symmetry, so that these two modes are doubly degenerate.

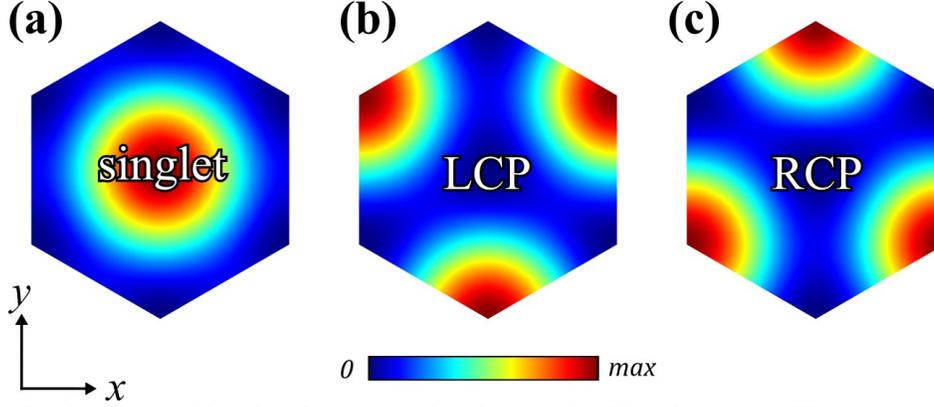

**Figure S2: Orthonormal basis of vacuum lattice at the $K$ point**. (a-c) The intensity plots of the singlet: $\left|H_z^{(K,\text{singlet})}\right|^2$, and the doublets: $\left|H_z^{(K,\text{LCP})}\right|^2$ and $\left|H_z^{(K,\text{RCP})}\right|^2$, respectively.

The unique phase of the amplitudes of RCP and LCP is responsible for the local rotating fields as shown in Fig.1(c). This symmetry property allows us to give an operational definition of the RCP and LCP fields without doing any plane-wave expansion on a more realistic photonic structure. As discussed in the main texts, rotating the field of RCP and LCP mode by $2\pi/3$ along z-axis bring the field back to itself with merely a phase factor:

$$\mathcal{R}_3 H_z^{(K,\text{RCP})} = \eta^* H_z^{(K,\text{RCP})} \tag{S2.1}$$

$$\mathcal{R}_3 H_z^{(K,\text{LCP})} = \eta H_z^{(K,\text{LCP})} \tag{S2.2}$$

One can then use this simple symmetry property to demonstrate that a perturbation like Fig.1(b) with $C_3$ wave vector symmetry does not couple RCP and LCP, so that its corresponding perturbation matrix are diagonalized in the circularly-polarized (CP) basis. To prove this, the cross overlap integrals between the RCP and LCP is calculated [ 1]:

$$\Delta_{P,K}^{RL} = -\int_V \Delta\epsilon(\mathbf{r}_\perp) \cdot (\mathbf{e}_\perp^{R*} \cdot \mathbf{e}_\perp^L) \, dV \tag{S3}$$

where $\Delta V$ is the pertubated volume, and $\Delta\epsilon(\mathbf{r}_\perp) = \pm(\epsilon_{\text{Si}} - 1)$ is the changing permittivity after perturbation (circular to triangular rod). The $\pm$ sign of $\Delta\epsilon(\mathbf{r}_\perp)$ depends on whether the vacuum region is replaced by Si or vice versa. Because the system is unchanged under $\mathcal{R}_3$, the following equality must satisfied:

$$\Delta_{P,K}^{RL} = \mathcal{R}_3 \Delta_{P,K}^{RL} \tag{S4}$$

The explicit operation of $\mathcal{R}_3 \equiv \mathcal{R}_{\theta=2\pi/3}$ contents two parts: (i) mapping the argument of a function such that $(x, y, z) \to (x', y', z')$ where $(x', y', z')^T = \bar{\bar{R}}_{-\theta}(x, y, z)^T$; $\bar{\bar{R}}_\theta = \begin{pmatrix} \cos\theta & -\sin\theta & 0 \\ \sin\theta & \cos\theta & 0 \\ 0 & 0 & 1 \end{pmatrix}$ and $\theta = 2\pi/3$; (ii) rotating the components of a vector such that $(v_x, v_y, v_z)^T \to \bar{\bar{R}}_\theta(v_x, v_y, v_z)^T$. So together one has the rotation operation of a vector field:

$$\mathcal{R}_\theta \begin{pmatrix} v_x(x,y,z) \\ v_y(x,y,z) \\ v_z(x,y,z) \end{pmatrix} = \bar{\bar{R}}_\theta \begin{pmatrix} v_x(x',y',z') \\ v_y(x',y',z') \\ v_z(x',y',z') \end{pmatrix}$$

From the definition of the $R$ and $L$ fields, we know that $\mathcal{R}_3 \mathbf{e}_\perp^R = \eta^* \mathbf{e}_\perp^R$ and $\mathcal{R}_3 \mathbf{e}_\perp^L = \eta \mathbf{e}_\perp^L$. Thus, Eq.(S4) becomes

$$\Delta_{P,K}^{RL} = \int_{\Delta V} \Delta\epsilon(\boldsymbol{r}_\perp) \cdot [\mathcal{R}_3 \boldsymbol{e}_\perp^{R*} \cdot \mathcal{R}_3 \boldsymbol{e}_\perp^L] \, dV$$

$$= \eta^2 \int_{\Delta V} \Delta\epsilon(\boldsymbol{r}_\perp) \cdot [\boldsymbol{e}_\perp^{R*} \cdot \boldsymbol{e}_\perp^L] \, dV$$

$$= \eta^2 \Delta_{P,K}^{RL}$$

To satisfy the above equality, $\Delta_{P,K}^{RL}$ has to be zero. Similarly, $\Delta_{P,K}^{LR} = 0$. Therefore the perturbation matrix $\Delta_{P,K} \equiv \begin{pmatrix} \Delta_{P,K}^{RR} & \Delta_{P,K}^{RL} \\ \Delta_{P,K}^{LR} & \Delta_{P,K}^{LL} \end{pmatrix}$ and the corresponding perturbation Hamiltonian must have diagonalized form.

*Suppression of Inter-Valley Scattering and Conservation of the Valley Degree of Freedom under 'zigzag' Perturbation of the Interface between PTIs with Different Topological Valley Indices:*
In this section, we analytically explain under which type of perturbations the inter-valley scattering is suppressed. In other words, once an edge wave is launched at certain valley on an interface, while it is propagating, the scattering only takes place within the valley in an effective 2D BZ of the whole system. Although the system such as Figs.(3,4) are not, strictly speaking, periodic structures, one can imagine that the triangular-rod perturbations are so weak that the evanescent field of a TPEW has very large decay length in the transverse direction. This TPEW is then essentially the same bulk propagating mode near the Dirac point of an unperturbed photonic graphene. The bulk mode propagating along the same direction of the TPEW is made of the superposition of RCP and LCP modes: $H_z(\boldsymbol{r}_\perp) = H_z^{(K,\text{RCP})}(\boldsymbol{r}_\perp) + e^{i\theta} H_z^{(K,\text{LCP})}(\boldsymbol{r}_\perp)$, where $\theta = \tan^{-1}(k_y/k_x)$. As one increases the perturbations, the evanescent field of the TPEW becomes more localized to the interface; however the TPEW is still, following this reasoning, built up from the basic RCP and LCP modes of unperturbed photonic graphene. Thus to understand the scattering properties between the edge states along different edges, studying the field overlaps of the unperturbed photonic graphene is sufficient as long as the PTIs are still in perturbation regime.

Because the valley conservation in our discussion does not involve spin DOF, we can consider only one representative field component of the eigenmodes, say $H_z(\boldsymbol{r}_\perp) = \psi_{K(K')}(\boldsymbol{r})$ for the TE modes, where $\boldsymbol{r} = (x, y)$. At Dirac points of the $K$ and $K'$ valleys, $\psi_{K(K')}(\boldsymbol{r})$ can be expressed in the Bloch form:

$$\psi_K(\boldsymbol{r}) = u_K(\boldsymbol{r}) e^{i\boldsymbol{K}\cdot\boldsymbol{r}} \tag{S5}$$

$$\psi_{K'}(\boldsymbol{r}) = u_{K'}(\boldsymbol{r}) e^{i\boldsymbol{K'}\cdot\boldsymbol{r}}, \tag{S6}$$

where $u_K(\boldsymbol{r})$ and $u_{K'}(\boldsymbol{r})$ are functions with the periodicity of the lattice; $\boldsymbol{K} = \boldsymbol{e}_x \, 4\pi/3a_0$ and $\boldsymbol{K'} = -\boldsymbol{e}_x \, 4\pi/3a_0$; $\boldsymbol{r} = (x, y)$. The overlapped field of the same valley (intra-valley), the $K$ valley, is then

$$\psi_K^* \psi_K = u_K^* u_K e^{i(\boldsymbol{K}-\boldsymbol{K})\cdot\boldsymbol{r}} = u_K^* u_K. \tag{S7}$$

Since $u_K^* u_K$ has the periodicity identical to the lattice, one can expect that the perturbation sitting on some lattice sites (with the same finite volume at the center of a unit cell) are always going to add up. This is because the value of the overlap integral of $\psi_K^* \psi_K$ over perturbation volume in every unit cell is exactly the same. However this is not the case for the inter-valley (between two different valleys) one. We shall see that for the inter-valley overlapped field $\psi_{K'}^* \psi_K$, the perturbations on the nearby sites tend to cancel with each other. The inter-valley overlapped field has in fact different periodicity from that of the lattice:

$$\psi_{K'}^* \psi_K = e^{i(K-K')\cdot r} u_{K'}^* u_K$$

$$= e^{i\frac{2}{3}(b_1+b_2)\cdot r} \sum_{m,n} a_{mn}^{KK'} e^{i(mb_1+nb_2)\cdot r}$$

$$= \sum_{m,n} a_{mn}^{KK'} e^{i[(3m+2)b_1' + (3n+2)b_2']\cdot r}$$

$$= \sum_{m',n'} b_{m'n'}^{KK'} e^{i(m'b_1' + n'b_2')\cdot r}$$

(S8)

where $b_{1,2}$ is the reciprocal lattice vectors, and $b_{1,2}' = 1/3\, b_{1,2}$, $m' = 3m+2$, $n' = 3n+2$. The last line of Eq.(S8) shows that $\psi_{K'}^* \psi_K$ has the original hexagonal symmetry, but has the period changed to $3a_0$ characterized by a new set of reciprocal vectors $b_1'$ and $b_2'$. With this mental picture of the inter-valley overlapped fields, we derive the special condition of the perturbation that gives zero overlap integral of $\psi_{K'}^* \psi_K$ (i.e. the perturbation that conserves the valley DOF).

Consider again the perturbations sitting on some lattice sites with the same finite volume. The difference, this time, is that the overlap integral corresponding to each site is no longer identical. The value of the integral varies from site to site with period $3a_0$ along 6 special directions ($l\pi/6$ with $l = 0,1,\ldots,5$) which is known as the directions of zigzag edge or the direction of the $K$ and $K'$ points. If we place the perturbations on 3 lattice sites in series along the directions of zigzag edge, the overlap integral reads

$$\int_{\Delta V = \Delta V_1 + \Delta V_2 + \Delta V_3} dV\, \psi_{K'}^* \psi_K$$

$$= \int_{\lambda=0}^{3a_0} \int_{\lambda_\perp = -\infty}^{\infty} d\lambda\, d\lambda_\perp\, w(\lambda_\perp)\, h \sum_{l'} \Lambda_{l'} e^{il'\frac{2\pi}{a_0}\hat{\lambda}\cdot r} \sum_{m',n'} b_{m'n'}^{KK'} e^{i(m'b_1' + n'b_2')\cdot r}$$

$$= \Delta A \sum_{l',m',n'} \Lambda_{l'} b_{m'n'}^{KK'} \int_{\lambda=0}^{3a_0} d\lambda\, e^{i\left(l'\frac{2\pi}{a_0}\hat{\lambda} + m'b_1' + n'b_2'\right)\cdot r}$$

$$= \Delta A \sum_{l',m',n'} \Lambda_{l'} b_{m'n'}^{KK'} \frac{3ia_0}{2\pi} \cdot \frac{1 - e^{i2\pi[3l' + (m'+n')\cos\phi + 1/\sqrt{3}(m'-n')\sin\phi]}}{3l' + (m'+n')\cos\phi + 1/\sqrt{3}(m'-n')\sin\phi}$$

$$= \Delta A \sum_{l',m',n'} \Lambda_{l'} b_{m'n'}^{KK'} \frac{3ia_0 (1 - e^{i2\pi I})}{2\pi I} = 0$$

(S9)

where $\hat{\lambda} = [\cos\phi, \sin\phi]$ and $r = [\lambda\cos\phi, \lambda\sin\phi]$ with $\phi = l\pi/6$ and $l = 0,1,\ldots,5$; $w(\lambda_\perp)$: a localized function along $\lambda_\perp$ (the coordinate along the perpendicular direction of $\hat{\lambda}$), $h$: the height of the perturbation volume, and the Fourier series in the direction of $\hat{\lambda}$ make the integration region continuous; $\Delta A = \int_{\lambda_\perp = -\infty}^{\infty} d\lambda_\perp\, w(\lambda_\perp)\, h$ is the vertical cross section of the perturbation volume. The integer I in the last line of Eq. (S16) is

$$I \equiv 3l' + (m' + n')\cos\phi + 1/\sqrt{3}(m' - n')\sin\phi = \begin{cases} 3l' \pm m' \pm n', & l = 0, 3 \\ 3l' \pm m', & l = 1, 4 \\ 3l' \mp n', & l = 2, 5 \end{cases} \quad (S10)$$

As shown in Eq. (S10), I is an integer as long as $\phi$ is of that along the zigzag edge, and if so, the overlap integral [Eq. (S9)] is identically zero. This result shows that as long as the perturbations are designed to be the same at wherever lattice sites we put them and they perturb $3N$ lattice sites in series along the direction of zigzag edge, the inter-valley scattering is prohibited and the valley DOF is conserved. We refer this type of perturbations 'zigzag' as opposed to the other 'armchair' type of perturbation.

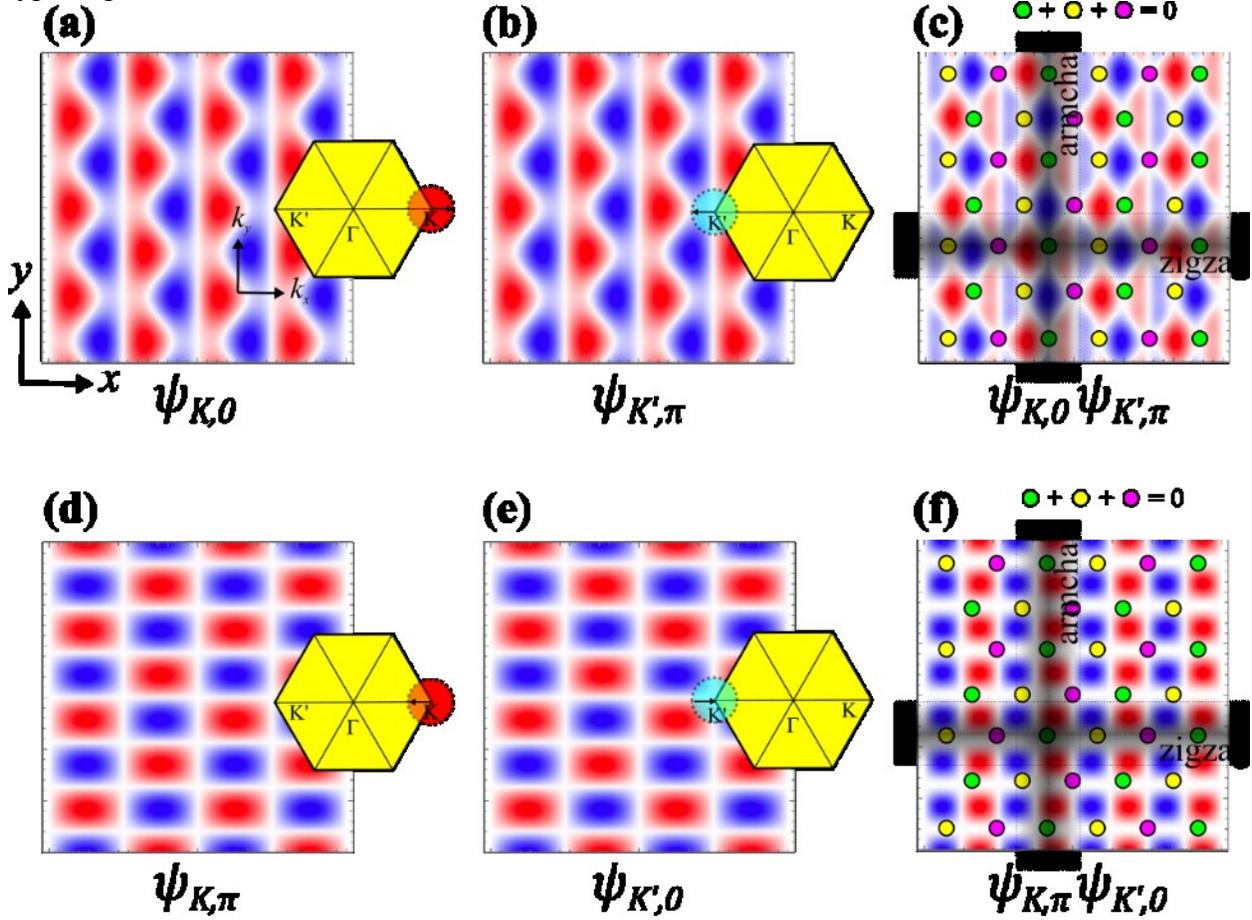

**Figure S3: Inter-valley overlap fields and the valley conservation**. **(a,b)** Field profiles of forward-moving $K$-valley and backward-moving $K'$-valley respectively. **(c)** The overlap field between (a) and (b); color dots distinct the different strength of coupling proportional to the overlap integral in the region near the lattice sites; Gray bands mark the direction of zigzag and armchair. **(d,e,f)** Same as (a,b,c) but for the case of backward-moving $K$-valley and forward-moving $K'$-valley.

Fig.S3 illustrates the idea of the valley conservation under zigzag perturbation. Figs.S3(a-c) show the overlap field between forward-moving $K$-valley and backward-moving $K'$-valley, and Figs.S3(d-f) show the overlap field between backward-moving $K$-valley and forward-moving $K'$-valley. The color dots mark the different strength of on-site perturbations which are directly proportional to the field overlap integral in the perturbation region. One can see that along zigzag direction the perturbations of different lattice sites tend to cancel each other as we analytically shown above, whereas they tend to add up along the armchair direction. That is, the armchair-type of perturbation in general do not conserve the valley DOF.

*References*
1. Pozar, D. M., Microwave Engineering, 4th ed. (Wiley Global Education, 2011).